\documentclass[%
 reprint,
amsmath,amssymb,
prl,
]{revtex4-1}
\usepackage{graphicx}
\usepackage{dcolumn}
\usepackage{bm}
\usepackage{epstopdf}

\begin{document}

\preprint{APS/123-QED}

\title{Symmetry-unprotected nodes or gap minima in $s_{++}$ state of FeSe single crystal}

\author{Yue Sun,$^{1,2}$}
\email{sunyue@issp.u-tokyo.ac.jp}
\author{Akiyoshi Park,$^1$ Sunseng Pyon,$^1$ Tsuyoshi Tamegai,$^1$ and Hisashi Kitamura$^3$}

\affiliation{%
$^1$Department of Applied Physics, The University of Tokyo, Bunkyo-ku, Tokyo 113-8656, Japan\\
$^2$Institute for Solid State Physics (ISSP), The University of Tokyo, Kashiwa, Chiba 277-8581, Japan\\
$^3$National Institute of Radiological Sciences, National Institutes for Quantum and Radiological Science and Technology, Chiba, 263-8555, Japan}


\begin{abstract}
We report the study on superconducting pairing mechanism of FeSe via the pair-breaking effects induced by H$^+$-irradiation combined with low-temperature specific heat measurements. A multi-gap structure with nodes or gap minima is suggested in a clean FeSe by the specific heat results. The suppression of critical temperature $T_c$ with increasing the defect density manifests a two-step behavior. When the increase in the residual resistivity is small, $\Delta\rho_0$ $<$ $\sim$4.5 $\mu\Omega$cm, $T_c$ is gradually suppressed with increasing the density of scattering centers, suggesting the presence of symmetry-unprotected nodes or gap minima. However, for $\Delta\rho_0$ $>$ $\sim$4.5 $\mu\Omega$cm, $T_c$ is almost independent of the scattering, which indicates that the nodes or gap minima are lifted and the order parameter becomes almost isotropic without sign change. Thus, the superconductivity in FeSe is found to be realized in symmetry-unprotected nodal or highly anisotropic $s_{++}$ state.

\end{abstract}

\maketitle
FeSe has attracted much attention because it is an intriguing candidate for both searching for high-temperature superconductivity (SC) and probing the superconducting mechanism. Although the initial $T_c$ in FeSe is below 10 K \cite{HsuFongChiFeSediscovery}, it can be easily enhanced up to 37 K under pressure \cite{MedvedevNatMat} and over 40 K by intercalating spacer layers \cite{BurrardNatMat}. Recently, the monolayer of FeSe grown on SrTiO$_3$ even shows a sign of $T_c$ over 100 K \cite{GeNatMatter}. It undergoes only the structural transition from tetragonal to orthorhombic at $T_s$ $\sim$87 K without long-range magnetic order at any temperature \cite{McQueenPRL}. Such unique features make FeSe an ideal material to study the nematic state, which is often referred as the origin of structural transition and could be related directly to high-temperature SC \cite{FernandesNatPhy}. The Fermi energy ($E_F$) of FeSe is found to be extremely small and comparable to the superconducting energy gap ($\Delta$), indicating that superconductivity in FeSe is realized in the crossover regime from Bardeen-Cooper-Schrieffer (BCS) to Bose-Einstein-condensation (BEC) \cite{Kasahara18112014}.

The superconducting gap structure is crucial to the understanding of these intriguing properties and the unexpected high $T_c$ in FeSe system. Kasahara $et$ $al$. \cite{Kasahara18112014} reported nodes in the gap structure of FeSe bulk single crystal based on the V-shaped spectrum observed in scanning tunneling spectroscopy, a nearly linearly temperature dependent penetration depth at low temperatures, and a large residual thermal conductivity. However, the nodeless gap structure in bulk FeSe was supported by the low-temperature specific heat \cite{LinFeSeSHPRB,LinJiaoarxiv}, lower critical field \cite{AbdelHc1FeSePRB}, and thermal conductivity measurements reported by other groups \cite{FeSeoldthermalPRB,hopearxiv}. Despite the controversy on the gap structure, there are few studies on bulk FeSe to distinguish the inter-band sign-reversed $s_\pm$ state (SC mediated by antiferromagnetic spin fluctuations between different bands with oppsite sign) \cite{MazinS} and the sign-preserving $s_{++}$ state  (SC mediated by orbital fluctuations between different bands with the same sign) \cite{KontaniPRL}. Actually, distinguishing between the $s_\pm$ and $s_{++}$ states in iron-based superconductors (IBSs) is always a challenging task since a phase-sensitive probe is necessary. It is even more difficult in the case of FeSe because of the interference from the possible existence of gap nodes.

The nonmagnetic disorder induced by light-particle irradiations, like the electron and H$^+$ have been proved to be an effective method to identify the gap structures of superconductors \cite{AlloulRevModPhys.81.45}. Based on the Anderson's theorem, $T_c$ of the conventional BCS superconductor with an isotropic gap is robust against nonmagnetic impurities \cite{Anderson}. In the case of superconductors with symmetry protected nodes such as $d$-wave, fast and continuous suppression of $T_c$ is expected, which has already been proved by experiments in YBa$_2$Cu$_3$O$_{7-\delta}$ \cite{AlloulRevModPhys.81.45,RullierPRL}. By contrast, in the case of SC with symmetry-unprotected nodes, such as the nodal $s$-wave, the nodes can be lifted by certain amount of scattering centers, which means that the symmetry unprotected nodal $s$-wave can be tuned into nodeless $s$-wave by the introduction of scattering centers \cite{WangPRBdisorder,MizukamiNatcom}. Even the $s_\pm$ and $s_{++}$ states can be distinguished by the nonmagnetic impurity effect based on the difference in the $T_c$ suppression rate \cite{OnariPRLdisorder,WangPRBdisorder,ProzorovPRXBaPirra}. Thus, the nonmagnetic disorder effect can be a unique and promising way to probe the pairing mechanism of FeSe.

In this Rapid Communication, we report on the effects of nonmagnetic scatterings in a clean FeSe single crystal produced by H$^+$-irradiation combined with low-temperature specific heat measurements. The results suggest that the SC in a clean FeSe is $s_{++}$ state with symmetry-unprotected nodes or gap minima.

\begin{figure*}\center
\includegraphics[width=16cm]{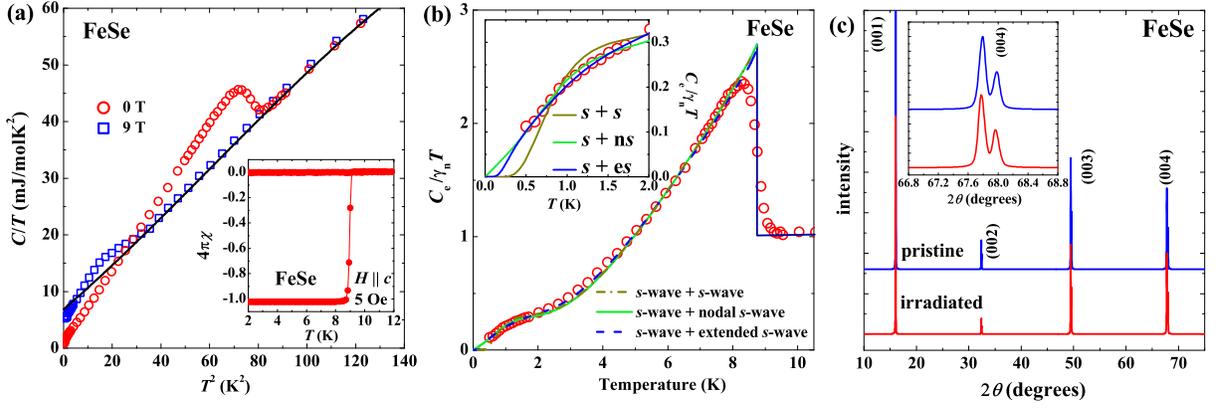}\\
\caption{(a) Specific heat divided by temperature, $C/T$ vs $T^2$, measured under 0 and 9 T. The solid line represents the fit to the normal state specific heat. Inset is the temperature dependence of magnetic susceptibility $\chi$. (b) Normalized zero field electronic specific heat $C_e/\gamma_nT$ vs $T$ together with the fit lines of $s$ + $s$-wave ($2\Delta_L/k_BT_c=4.35$, $2\Delta_S/k_BT_c=0.74$, $\gamma_1=0.74$, $\gamma_2=0.26$), $s$ + nodal $s$-wave ($2\Delta_s/k_BT_c=4.35$, $2\Delta_{ns}/k_BT_c=1.06$, $\gamma_1=0.74$, $\gamma_2=0.26$) and $s$ + extended $s$-wave ($2\Delta_s/k_BT_c=4.35$, $2\Delta_e/k_BT_c=0.85$, $\gamma_1=0.71$, $\gamma_2=0.29$, $\alpha=0.72$). Inset is the enlarged low temperature part. (c) XRD patterns for the FeSe before and after irradiating H$^+$ up to 5$\times$10$^{16}$ ions/cm$^2$. Inset is the enlarged part of the (004) peaks.}\label{}
\end{figure*}

High-quality FeSe single crystals were grown by the vapor transport method. Details of the crystal growth and sample quality have been reported in our previous publications \cite{SunPhysRevB.93.104502,SunYueAMRFeSe}. Single crystals used for the irradiation experiments were cleaved to thin plates with thickness $\sim$25 $\mu$m along the $c$-axis, which is much smaller than the projected range of 3-MeV H$^+$ for FeSe of $\sim$50 $\mu$m \cite{irradiationrange}. The crystal was loaded onto a sapphire plate with small amount of Apiezon grease, and cooled down by a closed-cycle refrigerator at the terminal of the irradiation port. There is no strain effects during the experiments as shown in Supplementary Fig. S1 \cite{supplement}. The 3-MeV H$^+$ was irradiated parallel to the $c$-axis of the crystal at 50 K avoiding the effect of thermal annealing. Resistivity measurements were performed $in$ $situ$ immediately after each irradiation by a standard four-probe method. Thus, the crystal was not exposed to air or temperatures higher than 50 K, which avoided the possible annealing effect as explained in Supplementary Fig. S2 \cite{supplement}. Furthermore, there is no sample-dependent influence since all the transport measurements were done on the same piece of single crystal. The H$^+$-irradiation was performed at Heavy Ion Medical Accelerator in Chiba (HIMAC) in National Institute of Radiological Sciences (NIRS). Structure of the crystal was characterized by means of X-ray diffraction (XRD) with Cu-K$\alpha$ radiation. Magnetization measurements were performed using a commercial SQUID magnetometer (MPMS-XL5). Specific heat data were obtained using the Physical Property Measurement System (PPMS, Quantum Design) with $^3$He refrigerator.

Inset of Fig. 1(a) shows the temperature dependence of magnetic susceptibility $\chi$ for the pristine FeSe single crystal, which displays $T_c$ $\sim$ 9.0 K with a sharp transition width. The main panel of Fig. 1(a) shows the specific heat of FeSe divided by temperature $C/T$ as a function of $T^2$ under 0 and 9 T. A clear jump associated with superconducting transition is observed around 9 K under zero field, which is consistent with susceptibility measurement. The normal state specific heat can be fitted by the sum of electronic part and phononic part: $C_n/T$ = $\gamma_n$ + $\beta_3T^2$ + $\beta_5T^4$ + $\beta_7T^6$. The fitting result was shown as the solid line in Fig. 1(a) giving $\gamma_n$ = 6.86 mJ/mol$\cdot$K$^2$, $\beta_3$ = 0.37 mJ/mol$\cdot$K$^4$, $\beta_5$ = 0.001 mJ/mol$\cdot$K$^6$, and $\beta_7$ = -5.72 $\times$ 10$^{-6}$ mJ/mol$\cdot$K$^8$. The normalized specific heat jump at $T_c$, $\Delta C/\gamma_nT_c$ is estimated to be 1.62, which is larger than the weak-coupling value 1.43 of BCS theory, implying that the superconductivity in FeSe is in the strong-coupling. Zero-field electronic specific heat $C_e/T$ obtained from subtracting the phonon terms, is shown in Fig. 1(b). Other than the specific heat jump at $T_c$, a second drop at around 1.2 K is also observed. This is a typical behavior of a two-gap superconductor such as MgB$_2$ \cite{MgB2PhysRevLetttwogap}, suggesting that FeSe is not a single-gap superconductor.

To get more information about the gap structure, the zero-field electronic specific heat is fitted by the following formula based on the BCS theory,
\begin{equation}
\label{eq.1}
\begin{split}
C_e=2N(0)\beta k_B\frac{1}{4\pi}\int_0^{2\pi}d\phi\int_0^\pi d\theta\sin\theta \\
\times\int_{-\hbar\omega_D}^{\hbar\omega_D}(-\frac{\partial f}{\partial E})(E^2+\frac{1}{2}\beta\frac{d\Delta^2}{d\beta})d\varepsilon,
\end{split}
\end{equation}
where $N(0)$ is the density of states at the Fermi surface, $\beta$ = 1/$k_BT$, and $E = \sqrt{\varepsilon^2+\Delta^2}$. The order parameters used to fit the data are $\Delta$ = $\gamma_1\Delta_L$ + $\gamma_2\Delta_S$ for two isotropic $s$-wave ($s$ + $s$), $\Delta$ = $\gamma_1\Delta_s$ + $\gamma_2\Delta_{ns}\cos2\phi$ for an isotropic $s$-wave plus a nodal $s$-wave with two line nodes ($s$ + $ns$), and $\Delta$ = $\gamma_1\Delta_s$ + $\gamma_2\Delta_e(1+\alpha\cos2\phi)$ for an isotropic $s$-wave plus an extended $s$-wave (with gap minima, $\alpha$ denotes the gap anisotropy) ($s$ +  $es$). The enlarged low-temperature part shown in the inset of Fig. 1(b) manifests that the $s$ + nodal $s$-wave and $s$ + extended $s$-wave can fit the data better than the $s$ + $s$-wave, which indicates the existence of nodes or gap minima in FeSe.

Fig. 1(c) shows the single crystal XRD pattern measured at room temperature for the crystal before and after irradiating H$^+$ up to 5$\times$10$^{16}$ ions/cm$^2$. After the irradiation, the positions of (00$l$) peaks are almost the same as the pristine one, which can be seen more clearly in the enlarged part of (004) peaks shown in the inset of Fig. 1(c). Evidently, no obvious broadening of the XRD peaks can be observed after the irradiation. The split of the peaks are coming from the $K_{\alpha1}$ and $K_{\alpha2}$ lines of the Cu. To obtain the lattice constant $a$/$b$, we fix the 2$\theta$ (angle between incident light and scattered light) to the higher (00$l$) peaks, like (003) or (004), then scan the incident angle $\theta$ to let the (103) or (104) peaks meet the Ewald Sphere of the reciprocal space.  The lattice constants are estimated as $c$ = (5.524$\pm$0.002) {\AA}, $a$ = (3.777$\pm$0.002) {\AA} for the pristine, and $c$ = (5.524$\pm$0.002) {\AA}, $a$ = (3.774$\pm$0.002) {\AA} for the irradiated crystals. We also confirm that temperature dependencies of Hall coefficients for the pristine and irradiated crystals are almost identical except for temperatures below 20 K as shown in Supplementary Fig. S3 \cite{supplement}. If we ignore possible annealing effects during the XRD and Hall coefficient measurements, we can conclude that the H$^+$-irradiation has little effects on both crystal and electronic band structures.

Fig. 2(a) shows the temperature dependence of resistivity for FeSe irradiated with increasing amount of H$^+$. The $\rho$-$T$ curves in the normal state show a parallel shift upon increasing the irradiation dose without obvious upturn at low temperatures, and the superconducting transition width shows no significant broadening. These facts indicate that the introduced point defects are nonmagnetic with no trace of localization effects. The residual resistivity $\rho_0$ was obtained by linearly extrapolating $\rho$-$T$ curves in the normal state above $T_c$ to $T$ = 0 K as shown by the dashed lines in Fig. 2(a). The value of $\rho_0$ for the pristine crystal is $\sim$ 1 $\mu\Omega$ cm, which again confirms the very high quality of the crystal used in this study.

The evolution of the difference of the residual resistivity between the irradiated and pristine ones, namely,  $\Delta\rho_0$ = $\rho_0^{irr}$ - $\rho_0^{unirr}$ with the dose of irradiated H$^+$ are shown in the Fig. 2(b). An almost linear increase in $\Delta\rho_0$ is evident, which guarantees that the H$^+$-irradiation introduces defects systematically. On the other hand, the value of $T_c$ is obviously suppressed after the irradiation. To avoid the possible ambiguity from the criteria of $T_c$, we obtained both $T_c^{onset}$ and $T_c^{mid}$ by the onset and the midpoint of the resistive transition as shown in the Supplementary Fig. S2. The evolutions of $T_c$ with the dose are also shown in Fig. 2(b). The values of $T_c$ are quickly suppressed in the small dose region, while maintaining almost dose-independent behavior when irradiated over 5$\times$10$^{15}$ ions/cm$^2$. Such behavior is quite different from those observed in iron pnictides, in which the $T_c$ is linearly suppressed with increasing the dose of irradiated particles although the suppression rate is sample dependent \cite{MizukamiNatcom,ProzorovPRXBaPirra,NakajimaPRBBaCoirra,TaenPRBBaKirr}.

\begin{figure}\center
\includegraphics[width=8.5cm]{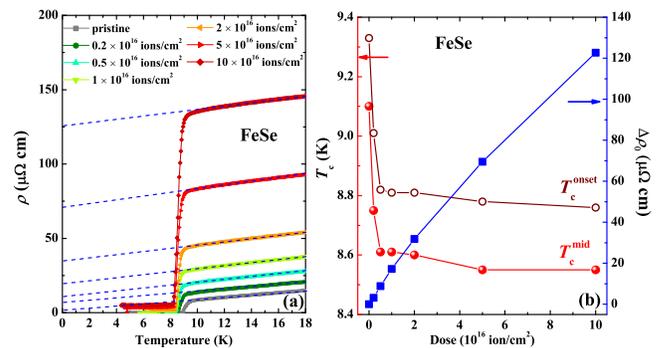}\\
\caption{(a) Temperature dependence of the resistivity for FeSe irradiated by H$^+$ with the dose of 0, 0.2, 0.5, 1, 2, 5, and 10 $\times$ 10$^{16}$ ions/cm$^2$. Dashed lines are linear extrapolations to zero temperature for estimating the residual resistivity $\rho_0$. (b) Dose dependence of $T_c^{onset}$ and $T_c^{mid}$, and the difference of the residual resistivity before and after irradiation ($\Delta\rho_0$).}\label{}
\end{figure}

For quantitative discussion on the pair-breaking effects by non-magnetic scatterings, we plot $T_c$ and $T_c/T_{c0}$ (inset) as a function of $\Delta\rho_0$ in Fig. 3(a). Obviously, the evolution of $T_c$ with $\Delta\rho_0$ manifests a two-step behavior. $\Delta\rho_0^*$ defined as the crossover value of $\Delta\rho_0$ separating the two regions, is $\sim$4.5 $\mu\Omega$cm  as pointed out by the arrow. For $\Delta\rho_0 < \Delta\rho_0^*$, $T_c$ is gradually suppressed with increasing $\Delta\rho_0$ at a slope of $\sim$100 K/m$\Omega$cm. On the other hand, for $\Delta\rho_0 > \Delta\rho_0^*$, $T_c$ is almost independent of $\Delta\rho_0$ with a negligible suppression of $T_c$ less than 1\% over 100 $\mu\Omega$cm.

In the case of conventional BCS superconductors, it has been well established that the nonmagnetic impurities have no influence to the superconducting gap or the $T_c$ due to Anderson's theorem \cite{Anderson}. By contrast, in superconductors with sign-reversed order parameter as in the case of $s_\pm$-wave with inter-band scattering, and symmetry-protected nodes such as the $d$-wave, an appreciable suppression of $T_c$ is observed. However, in both cases of $s_\pm$ and symmetry-protected nodes,  $T_c$ will continuously decrease with increasing $\Delta\rho_0$ as proposed by theoretical calculation and already confirmed in cuprates and iron pnictides \cite{AlloulRevModPhys.81.45,RullierPRL,OnariPRLdisorder,WangPRBdisorder,ProzorovPRXBaPirra,MizukamiNatcom,NakajimaPRBBaCoirra,TaenPRBBaKirr}. It is obviously different from the observation here in FeSe that the suppression only exists in the narrow region of $\Delta\rho_0$ $<$ $\Delta\rho_0^*$, and the value of $T_c$ is only suppressed by $\sim$0.5 K.

Based on the low-temperature specific heat results above, the nodes or gap minima are present in the pristine FeSe. It should be noted that even if nodes exist, they are not symmetry protected as proposed by the theoretical calculation  \cite{KreiselPRB} since they are absent in crystals with low quality \cite{FeSeoldthermalPRB}. When the nodal positions are not symmetry protected, as in the case of nodal $s$-wave, the accidental nodes can be lifted and the low-energy quasi-particle excitations are eliminated at a certain concentration of nonmagnetic scatterings \cite{WangPRBdisorder,MizukamiNatcom}. The theoretical calculation of the disorder effect based on a simple two-band model (FeSe is reported containing one hole-typed and one or two electron-typed bands at temperatures below $T_s$ \cite{WatsonPRB92}.) shows that the density of states at  the Fermi-level for the accidental nodes in $s$-wave are reduced to zero when the $\Delta\rho_0$ is only $\sim$4 $\mu\Omega$ cm \cite{WangPRBdisorder}. Although the calculation is performed on iron-pnictides, the lifting of the accidental nodes by a small value of $\Delta\rho_0$ is confirmed, and the value of $\Delta\rho_0^*$ is found independent of the ratio of inter-band to intra-band scattering. Another explanation of the initial fast $T_c$ suppression is the averaging effect of highly anisotropic gap by disorder, which increases the gap minima, and reduces the anisotropy. The gap minima emerge when the nodes are lifted but the anisotropy is not averaged \cite{hopearxiv}, which will be discussed in detail later.
\begin{figure}\center
\includegraphics[width=7.5cm]{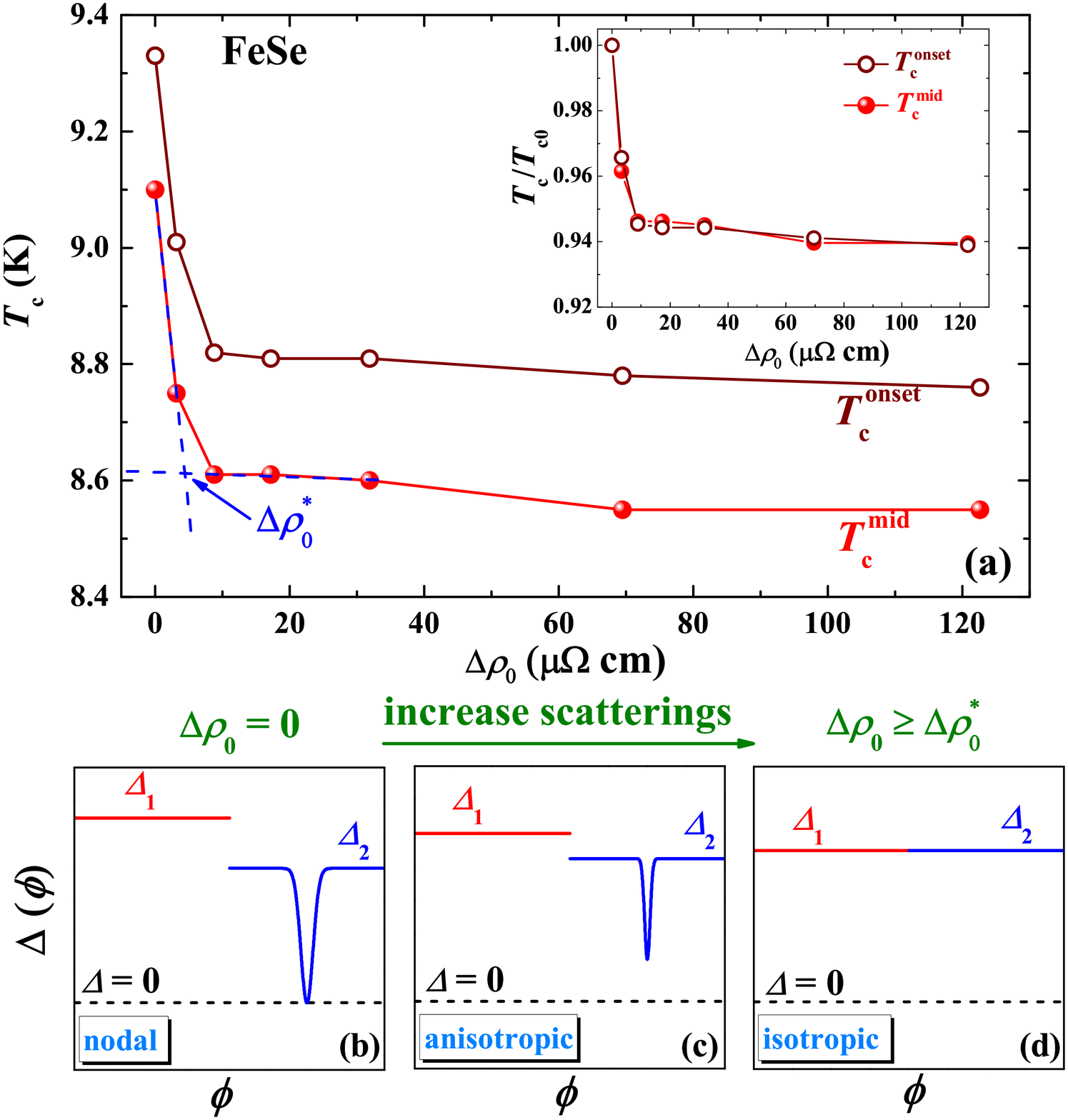}\\
\caption{(a) $T_c^{onset}$ and $T_c^{mid}$ as a function of $\Delta\rho_0$ for FeSe. The dashed lines show the definition of crossover point of $\Delta\rho_0^*$ by linear extrapolating the $T_c$ at two steps. Inset is the normalized $T_c/T_{c0}$ as a function of $\Delta\rho_0$. (b)-(d) Schematic evolution of order parameters as the density of scattering centers is increased.}\label{}
\end{figure}

After the nodes or the gap minima are totally lifted or averaged by the introduced scatterings, the wave function of the fully-gapped state in IBSs has two possibilities: the sign-reversed $s_\pm$ state \cite{MazinS} and the sign-preserved $s_{++}$ state \cite{KontaniPRL}, and both cases can be either isotropic or anisotropic. In the case of $s_\pm$ state, the $T_c$ will be continuously suppressed with further introduction of scatterings in spite of the anisotropy, unless the inter-band scattering is absent, although the $T_c$ suppression rate depends on the ratio of inter-band to intra-band scatterings \cite{WangPRBdisorder}. In practice, the inter-band scattering has already been found to be crucial in the IBSs. Thus, for the IBSs with  $s_\pm$ state, the $T_c$ will be sensitive to the non-magnetic scatterings rather than scattering-independent as experimentally observed in BaFe$_2$(As$_{1-x}$P$_x$)$_2$ \cite{MizukamiNatcom}. Obviously, the nodal $s_\pm$ state is not compatible with the scattering-independent second step observed in FeSe. The anisotropic $s_{++}$ state can be also excluded since the value of $T_c$ will be continually suppressed by increasing the scattering centers because of the inter-band scattering. The only possible scenario is the isotropic sign-preserved $s_{++}$ state, in which the $T_c$ is almost unaffected by the scattering centers as proposed by the theoretical calculations \cite{OnariPRLdisorder,WangPRBdisorder}. Thus, the observed scattering-independent second step in FeSe is attributed to the isotropic $s_{++}$ state.

To clearly show the change in the gap function with scattering centers, evolution of the order parameter $\Delta$ versus azimuthal angle $\phi$ is schematically shown in Figs. 3(b)-(d). Here, we only consider a simple two-band situation, where one is fully gapped, while the other has nodes or gap minima. As the density of scattering centers increases, the gap structure of FeSe changes from (b) nodal or (c) anisotropic nodeless with gap minima to (d) isotropic nodeless $s_{++}$ state. For both the nodal and anisotropic cases, the $T_c$ will be suppressed with increasing scattering centers. Furthermore, since the suppression of $T_c$ only exists in a very small region, and the value of $T_c$ is only suppressed $\sim$0.5 K, the nodes or gap minima should be very narrow and easily lifted as shown in Figs. 3(b) and (c). Otherwise, the anisotropy in the gap cannot be smeared with the scattering that fast. In the present experiment, we cannot directly distinguish the nodal gap (Fig. 3(b)) from deep minima in anisotropic nodeless gap (Fig. 3(c)). Comparing the sample quality to that with nodes reported in Ref \cite{Kasahara18112014}, the single crystal used in the present experiment manifests slightly better quality (judging from the sharper SC transition in susceptibility, and smaller residual resistivity). Hence, SC in our pristine crystal may be in nodal state as shown in Fig. 3(b). The SC in a pure FeSe may be symmetry-unprotected nodal state, and the nodes are very sensitive to the disorder, which can be easily tuned into anisotropic nodeless, and finally to isotropic nodeless $s_{++}$ state by small amount of disorder. The reported controversy in the gap structure of bulk FeSe can be explained by the difference in sample quality, i.e. the amount of scattering centers formed in crystal growth.

The sign-preserved $s_{++}$ state is also proposed for the single layered FeSe on SrTiO$_3$ \cite{FanNatPhyMonoFeSe}, which is consistent with our observation in the bulk FeSe. Recently, an unexpected enhancement of $T_c$ about 0.4 K in FeSe after electron irradiation was reported by Teknowijoyo $et$ $al$. \cite{Teknowijoyoarxiv}, and explained by the local strengthening of the pairing interaction by irradiation-induced Frenkel pairs (between vacancies and interstitials). Compared to the electron, H$^+$ has similar particle energy but much larger mass, which will expel the irradiated points much far away from the crystal rather than into the interstitial sites \cite{MizukamiNatcom}. Hence, the Frenkel pairs are not the main defects in our H$^+$-irradiation experiment. The H$^+$-irradiation usually introduces point-like defects, while randomly distributed small clusters may be present when the irradiated dose is increased over a certain value \cite{SmyliePhysRevB}. More details about the defects are discussed in Supplementary S4 \cite{supplement}.  Since the crystals in Ref. \cite{Teknowijoyoarxiv} have been exposed to higher temperatures after irradiation, the density of twin boundaries and domains is supposed to be changed when the crystal was cooled again crossing the $T_s$ because more defects are present. The twin boundaries and the domains are found to affect the gap value and structure of FeSe \cite{WatashigePRX}, which may be another explanation of the enhancement of $T_c$. Actually, the enhanced value of $T_c$ after electron irradiation in Ref. \cite{Teknowijoyoarxiv} is just similar to that of our pristine crystal.

In summary, we studied the gap structure of FeSe via low-temperature specific heat and the pair-breaking effect. The multi-gap structure with nodes or gap minima is suggested by the specific heat results. The suppression of $T_c$ with increasing the scattering manifests obvious two-step behavior, which indicates that SC in a clean FeSe is $s_{++}$ state with symmetry-unprotected nodes or gap minima. The gap function of FeSe can be quickly tuned by a small amount of scattering centers, and evolves from nodal to anisotropic nodeless, and finally to isotropic nodeless $s_{++}$ state.

\emph{Note added}.$-$We notice a report of the $s_\pm$ pairing in FeSe by the Bogoliubov quasiparticle interference imaging measurements was published recently \cite{BQPIScience}.


\bibliography{protonirrreferences}

\begin{thebibliography}{35}%
\makeatletter
\providecommand \@ifxundefined [1]{%
 \@ifx{#1\undefined}
}%
\providecommand \@ifnum [1]{%
 \ifnum #1\expandafter \@firstoftwo
 \else \expandafter \@secondoftwo
 \fi
}%
\providecommand \@ifx [1]{%
 \ifx #1\expandafter \@firstoftwo
 \else \expandafter \@secondoftwo
 \fi
}%
\providecommand \natexlab [1]{#1}%
\providecommand \enquote  [1]{``#1''}%
\providecommand \bibnamefont  [1]{#1}%
\providecommand \bibfnamefont [1]{#1}%
\providecommand \citenamefont [1]{#1}%
\providecommand \href@noop [0]{\@secondoftwo}%
\providecommand \href [0]{\begingroup \@sanitize@url \@href}%
\providecommand \@href[1]{\@@startlink{#1}\@@href}%
\providecommand \@@href[1]{\endgroup#1\@@endlink}%
\providecommand \@sanitize@url [0]{\catcode `\\12\catcode `\$12\catcode
  `\&12\catcode `\#12\catcode `\^12\catcode `\_12\catcode `\%12\relax}%
\providecommand \@@startlink[1]{}%
\providecommand \@@endlink[0]{}%
\providecommand \url  [0]{\begingroup\@sanitize@url \@url }%
\providecommand \@url [1]{\endgroup\@href {#1}{\urlprefix }}%
\providecommand \urlprefix  [0]{URL }%
\providecommand \Eprint [0]{\href }%
\providecommand \doibase [0]{http://dx.doi.org/}%
\providecommand \selectlanguage [0]{\@gobble}%
\providecommand \bibinfo  [0]{\@secondoftwo}%
\providecommand \bibfield  [0]{\@secondoftwo}%
\providecommand \translation [1]{[#1]}%
\providecommand \BibitemOpen [0]{}%
\providecommand \bibitemStop [0]{}%
\providecommand \bibitemNoStop [0]{.\EOS\space}%
\providecommand \EOS [0]{\spacefactor3000\relax}%
\providecommand \BibitemShut  [1]{\csname bibitem#1\endcsname}%
\let\auto@bib@innerbib\@empty
\bibitem [{\citenamefont {Hsu}\ \emph {et~al.}(2008)\citenamefont {Hsu},
  \citenamefont {Luo}, \citenamefont {Yeh}, \citenamefont {Chen}, \citenamefont
  {Huang}, \citenamefont {Wu}, \citenamefont {Lee}, \citenamefont {Huang},
  \citenamefont {Chu}, \citenamefont {Yan},\ and\ \citenamefont
  {Wu}}]{HsuFongChiFeSediscovery}%
  \BibitemOpen
  \bibfield  {author} {\bibinfo {author} {\bibfnamefont {F.~C.}\ \bibnamefont
  {Hsu}}, \bibinfo {author} {\bibfnamefont {J.~Y.}\ \bibnamefont {Luo}},
  \bibinfo {author} {\bibfnamefont {K.~W.}\ \bibnamefont {Yeh}}, \bibinfo
  {author} {\bibfnamefont {T.~K.}\ \bibnamefont {Chen}}, \bibinfo {author}
  {\bibfnamefont {T.~W.}\ \bibnamefont {Huang}}, \bibinfo {author}
  {\bibfnamefont {P.~M.}\ \bibnamefont {Wu}}, \bibinfo {author} {\bibfnamefont
  {Y.~C.}\ \bibnamefont {Lee}}, \bibinfo {author} {\bibfnamefont {Y.-L.}\
  \bibnamefont {Huang}}, \bibinfo {author} {\bibfnamefont {Y.-Y.}\ \bibnamefont
  {Chu}}, \bibinfo {author} {\bibfnamefont {D.~C.}\ \bibnamefont {Yan}}, \ and\
  \bibinfo {author} {\bibfnamefont {M.~K.}\ \bibnamefont {Wu}},\ }\href@noop {}
  {\bibfield  {journal} {\bibinfo  {journal} {Proc. Nat. Acad. Sci.}\ }\textbf
  {\bibinfo {volume} {105}},\ \bibinfo {pages} {14262} (\bibinfo {year}
  {2008})}\BibitemShut {NoStop}%
\bibitem [{\citenamefont {Medvedev}\ \emph {et~al.}(2009)\citenamefont
  {Medvedev}, \citenamefont {McQueen}, \citenamefont {Troyan}, \citenamefont
  {Palasyuk}, \citenamefont {Eremets}, \citenamefont {Cava}, \citenamefont
  {Naghavi}, \citenamefont {Casper}, \citenamefont {Ksenofontov}, \citenamefont
  {Wortmann},\ and\ \citenamefont {Felser}}]{MedvedevNatMat}%
  \BibitemOpen
  \bibfield  {author} {\bibinfo {author} {\bibfnamefont {S.}~\bibnamefont
  {Medvedev}}, \bibinfo {author} {\bibfnamefont {T.~M.}\ \bibnamefont
  {McQueen}}, \bibinfo {author} {\bibfnamefont {I.~A.}\ \bibnamefont {Troyan}},
  \bibinfo {author} {\bibfnamefont {T.}~\bibnamefont {Palasyuk}}, \bibinfo
  {author} {\bibfnamefont {M.~I.}\ \bibnamefont {Eremets}}, \bibinfo {author}
  {\bibfnamefont {R.~J.}\ \bibnamefont {Cava}}, \bibinfo {author}
  {\bibfnamefont {S.}~\bibnamefont {Naghavi}}, \bibinfo {author} {\bibfnamefont
  {F.}~\bibnamefont {Casper}}, \bibinfo {author} {\bibfnamefont
  {V.}~\bibnamefont {Ksenofontov}}, \bibinfo {author} {\bibfnamefont
  {G.}~\bibnamefont {Wortmann}}, \ and\ \bibinfo {author} {\bibfnamefont
  {C.}~\bibnamefont {Felser}},\ }\href@noop {} {\bibfield  {journal} {\bibinfo
  {journal} {Nat. Mater.}\ }\textbf {\bibinfo {volume} {8}},\ \bibinfo {pages}
  {630} (\bibinfo {year} {2009})}\BibitemShut {NoStop}%
\bibitem [{\citenamefont {Burrard-Lucas}\ \emph {et~al.}(2013)\citenamefont
  {Burrard-Lucas}, \citenamefont {Free}, \citenamefont {Sedlmaier},
  \citenamefont {Wright}, \citenamefont {Cassidy}, \citenamefont {Hara},
  \citenamefont {Corkett}, \citenamefont {Lancaster}, \citenamefont {Baker},
  \citenamefont {Blundell},\ and\ \citenamefont {Clarke}}]{BurrardNatMat}%
  \BibitemOpen
  \bibfield  {author} {\bibinfo {author} {\bibfnamefont {M.}~\bibnamefont
  {Burrard-Lucas}}, \bibinfo {author} {\bibfnamefont {D.~G.}\ \bibnamefont
  {Free}}, \bibinfo {author} {\bibfnamefont {S.~J.}\ \bibnamefont {Sedlmaier}},
  \bibinfo {author} {\bibfnamefont {J.~D.}\ \bibnamefont {Wright}}, \bibinfo
  {author} {\bibfnamefont {S.~J.}\ \bibnamefont {Cassidy}}, \bibinfo {author}
  {\bibfnamefont {Y.}~\bibnamefont {Hara}}, \bibinfo {author} {\bibfnamefont
  {A.~J.}\ \bibnamefont {Corkett}}, \bibinfo {author} {\bibfnamefont
  {T.}~\bibnamefont {Lancaster}}, \bibinfo {author} {\bibfnamefont {P.~J.}\
  \bibnamefont {Baker}}, \bibinfo {author} {\bibfnamefont {S.~J.}\ \bibnamefont
  {Blundell}}, \ and\ \bibinfo {author} {\bibfnamefont {S.~J.}\ \bibnamefont
  {Clarke}},\ }\href@noop {} {\bibfield  {journal} {\bibinfo  {journal} {Nat.
  Mater.}\ }\textbf {\bibinfo {volume} {12}},\ \bibinfo {pages} {15} (\bibinfo
  {year} {2013})}\BibitemShut {NoStop}%
\bibitem [{\citenamefont {Ge}\ \emph {et~al.}(2015)\citenamefont {Ge},
  \citenamefont {Liu}, \citenamefont {Liu}, \citenamefont {Gao}, \citenamefont
  {Qian}, \citenamefont {Xue}, \citenamefont {Liu},\ and\ \citenamefont
  {Jia}}]{GeNatMatter}%
  \BibitemOpen
  \bibfield  {author} {\bibinfo {author} {\bibfnamefont {J.~F.}\ \bibnamefont
  {Ge}}, \bibinfo {author} {\bibfnamefont {Z.~L.}\ \bibnamefont {Liu}},
  \bibinfo {author} {\bibfnamefont {C.}~\bibnamefont {Liu}}, \bibinfo {author}
  {\bibfnamefont {C.~L.}\ \bibnamefont {Gao}}, \bibinfo {author} {\bibfnamefont
  {D.}~\bibnamefont {Qian}}, \bibinfo {author} {\bibfnamefont {Q.~K.}\
  \bibnamefont {Xue}}, \bibinfo {author} {\bibfnamefont {Y.}~\bibnamefont
  {Liu}}, \ and\ \bibinfo {author} {\bibfnamefont {J.~F.}\ \bibnamefont
  {Jia}},\ }\href@noop {} {\bibfield  {journal} {\bibinfo  {journal} {Nat
  Mater}\ }\textbf {\bibinfo {volume} {14}},\ \bibinfo {pages} {285} (\bibinfo
  {year} {2015})}\BibitemShut {NoStop}%
\bibitem [{\citenamefont {McQueen}\ \emph {et~al.}(2009)\citenamefont
  {McQueen}, \citenamefont {Williams}, \citenamefont {Stephens}, \citenamefont
  {Tao}, \citenamefont {Zhu}, \citenamefont {Ksenofontov}, \citenamefont
  {Casper}, \citenamefont {Felser},\ and\ \citenamefont {Cava}}]{McQueenPRL}%
  \BibitemOpen
  \bibfield  {author} {\bibinfo {author} {\bibfnamefont {T.~M.}\ \bibnamefont
  {McQueen}}, \bibinfo {author} {\bibfnamefont {A.~J.}\ \bibnamefont
  {Williams}}, \bibinfo {author} {\bibfnamefont {P.~W.}\ \bibnamefont
  {Stephens}}, \bibinfo {author} {\bibfnamefont {J.}~\bibnamefont {Tao}},
  \bibinfo {author} {\bibfnamefont {Y.}~\bibnamefont {Zhu}}, \bibinfo {author}
  {\bibfnamefont {V.}~\bibnamefont {Ksenofontov}}, \bibinfo {author}
  {\bibfnamefont {F.}~\bibnamefont {Casper}}, \bibinfo {author} {\bibfnamefont
  {C.}~\bibnamefont {Felser}}, \ and\ \bibinfo {author} {\bibfnamefont {R.~J.}\
  \bibnamefont {Cava}},\ }\href@noop {} {\bibfield  {journal} {\bibinfo
  {journal} {Phys. Rev. Lett.}\ }\textbf {\bibinfo {volume} {103}},\ \bibinfo
  {pages} {057002} (\bibinfo {year} {2009})}\BibitemShut {NoStop}%
\bibitem [{\citenamefont {Fernandes}\ \emph {et~al.}(2014)\citenamefont
  {Fernandes}, \citenamefont {Chubukov},\ and\ \citenamefont
  {Schmalian}}]{FernandesNatPhy}%
  \BibitemOpen
  \bibfield  {author} {\bibinfo {author} {\bibfnamefont {R.~M.}\ \bibnamefont
  {Fernandes}}, \bibinfo {author} {\bibfnamefont {A.~V.}\ \bibnamefont
  {Chubukov}}, \ and\ \bibinfo {author} {\bibfnamefont {J.}~\bibnamefont
  {Schmalian}},\ }\href@noop {} {\bibfield  {journal} {\bibinfo  {journal}
  {Nat. Phys.}\ }\textbf {\bibinfo {volume} {10}},\ \bibinfo {pages} {97}
  (\bibinfo {year} {2014})}\BibitemShut {NoStop}%
\bibitem [{\citenamefont {Kasahara}\ \emph {et~al.}(2014)\citenamefont
  {Kasahara}, \citenamefont {Watashige}, \citenamefont {Hanaguri},
  \citenamefont {Kohsaka}, \citenamefont {Yamashita}, \citenamefont
  {Shimoyama}, \citenamefont {Mizukami}, \citenamefont {Endo}, \citenamefont
  {Ikeda}, \citenamefont {Aoyama}, \citenamefont {Terashima}, \citenamefont
  {Uji}, \citenamefont {Wolf}, \citenamefont {von Löhneysen}, \citenamefont
  {Shibauchi},\ and\ \citenamefont {Matsuda}}]{Kasahara18112014}%
  \BibitemOpen
  \bibfield  {author} {\bibinfo {author} {\bibfnamefont {S.}~\bibnamefont
  {Kasahara}}, \bibinfo {author} {\bibfnamefont {T.}~\bibnamefont {Watashige}},
  \bibinfo {author} {\bibfnamefont {T.}~\bibnamefont {Hanaguri}}, \bibinfo
  {author} {\bibfnamefont {Y.}~\bibnamefont {Kohsaka}}, \bibinfo {author}
  {\bibfnamefont {T.}~\bibnamefont {Yamashita}}, \bibinfo {author}
  {\bibfnamefont {Y.}~\bibnamefont {Shimoyama}}, \bibinfo {author}
  {\bibfnamefont {Y.}~\bibnamefont {Mizukami}}, \bibinfo {author}
  {\bibfnamefont {R.}~\bibnamefont {Endo}}, \bibinfo {author} {\bibfnamefont
  {H.}~\bibnamefont {Ikeda}}, \bibinfo {author} {\bibfnamefont
  {K.}~\bibnamefont {Aoyama}}, \bibinfo {author} {\bibfnamefont
  {T.}~\bibnamefont {Terashima}}, \bibinfo {author} {\bibfnamefont
  {S.}~\bibnamefont {Uji}}, \bibinfo {author} {\bibfnamefont {T.}~\bibnamefont
  {Wolf}}, \bibinfo {author} {\bibfnamefont {H.}~\bibnamefont {von
  Löhneysen}}, \bibinfo {author} {\bibfnamefont {T.}~\bibnamefont
  {Shibauchi}}, \ and\ \bibinfo {author} {\bibfnamefont {Y.}~\bibnamefont
  {Matsuda}},\ }\href {\doibase 10.1073/pnas.1413477111} {\bibfield  {journal}
  {\bibinfo  {journal} {Proc. Nat. Acad. Sci.}\ }\textbf {\bibinfo {volume}
  {111}},\ \bibinfo {pages} {16309} (\bibinfo {year} {2014})}\BibitemShut
  {NoStop}%
\bibitem [{\citenamefont {Lin}\ \emph {et~al.}(2011)\citenamefont {Lin},
  \citenamefont {Hsieh}, \citenamefont {Chareev}, \citenamefont {Vasiliev},
  \citenamefont {Parsons},\ and\ \citenamefont {Yang}}]{LinFeSeSHPRB}%
  \BibitemOpen
  \bibfield  {author} {\bibinfo {author} {\bibfnamefont {J.~Y.}\ \bibnamefont
  {Lin}}, \bibinfo {author} {\bibfnamefont {Y.~S.}\ \bibnamefont {Hsieh}},
  \bibinfo {author} {\bibfnamefont {D.~A.}\ \bibnamefont {Chareev}}, \bibinfo
  {author} {\bibfnamefont {A.~N.}\ \bibnamefont {Vasiliev}}, \bibinfo {author}
  {\bibfnamefont {Y.}~\bibnamefont {Parsons}}, \ and\ \bibinfo {author}
  {\bibfnamefont {H.~D.}\ \bibnamefont {Yang}},\ }\href@noop {} {\bibfield
  {journal} {\bibinfo  {journal} {Phys. Rev. B}\ }\textbf {\bibinfo {volume}
  {84}},\ \bibinfo {pages} {220507} (\bibinfo {year} {2011})}\BibitemShut
  {NoStop}%
\bibitem [{\citenamefont {Lin}\ \emph {et~al.}(2016)\citenamefont {Lin},
  \citenamefont {Huang}, \citenamefont {R\"{o}{\ss}ler}, \citenamefont {Koz},
  \citenamefont {R\"{o}{\ss}ler}, \citenamefont {Schwarz},\ and\ \citenamefont
  {Wirth}}]{LinJiaoarxiv}%
  \BibitemOpen
  \bibfield  {author} {\bibinfo {author} {\bibfnamefont {J.}~\bibnamefont
  {Lin}}, \bibinfo {author} {\bibfnamefont {C.}~\bibnamefont {Huang}}, \bibinfo
  {author} {\bibfnamefont {S.}~\bibnamefont {R\"{o}{\ss}ler}}, \bibinfo
  {author} {\bibfnamefont {C.}~\bibnamefont {Koz}}, \bibinfo {author}
  {\bibfnamefont {U.~K.}\ \bibnamefont {R\"{o}{\ss}ler}}, \bibinfo {author}
  {\bibfnamefont {U.}~\bibnamefont {Schwarz}}, \ and\ \bibinfo {author}
  {\bibfnamefont {S.}~\bibnamefont {Wirth}},\ }\href@noop {} {\bibfield
  {journal} {\bibinfo  {journal} {arXiv:1605.01908}\ } (\bibinfo {year}
  {2016})}\BibitemShut {NoStop}%
\bibitem [{\citenamefont {Abdel-Hafiez}\ \emph {et~al.}(2013)\citenamefont
  {Abdel-Hafiez}, \citenamefont {Ge}, \citenamefont {Vasiliev}, \citenamefont
  {Chareev}, \citenamefont {Van~de Vondel}, \citenamefont {Moshchalkov},\ and\
  \citenamefont {Silhanek}}]{AbdelHc1FeSePRB}%
  \BibitemOpen
  \bibfield  {author} {\bibinfo {author} {\bibfnamefont {M.}~\bibnamefont
  {Abdel-Hafiez}}, \bibinfo {author} {\bibfnamefont {J.}~\bibnamefont {Ge}},
  \bibinfo {author} {\bibfnamefont {A.~N.}\ \bibnamefont {Vasiliev}}, \bibinfo
  {author} {\bibfnamefont {D.~A.}\ \bibnamefont {Chareev}}, \bibinfo {author}
  {\bibfnamefont {J.}~\bibnamefont {Van~de Vondel}}, \bibinfo {author}
  {\bibfnamefont {V.~V.}\ \bibnamefont {Moshchalkov}}, \ and\ \bibinfo {author}
  {\bibfnamefont {A.~V.}\ \bibnamefont {Silhanek}},\ }\href@noop {} {\bibfield
  {journal} {\bibinfo  {journal} {Phys. Rev. B}\ }\textbf {\bibinfo {volume}
  {88}},\ \bibinfo {pages} {174512} (\bibinfo {year} {2013})}\BibitemShut
  {NoStop}%
\bibitem [{\citenamefont {Dong}\ \emph {et~al.}(2009)\citenamefont {Dong},
  \citenamefont {Guan}, \citenamefont {Zhou}, \citenamefont {Qiu},
  \citenamefont {Ding}, \citenamefont {Zhang}, \citenamefont {Patel},
  \citenamefont {Xiao},\ and\ \citenamefont {Li}}]{FeSeoldthermalPRB}%
  \BibitemOpen
  \bibfield  {author} {\bibinfo {author} {\bibfnamefont {J.~K.}\ \bibnamefont
  {Dong}}, \bibinfo {author} {\bibfnamefont {T.~Y.}\ \bibnamefont {Guan}},
  \bibinfo {author} {\bibfnamefont {S.~Y.}\ \bibnamefont {Zhou}}, \bibinfo
  {author} {\bibfnamefont {X.}~\bibnamefont {Qiu}}, \bibinfo {author}
  {\bibfnamefont {L.}~\bibnamefont {Ding}}, \bibinfo {author} {\bibfnamefont
  {C.}~\bibnamefont {Zhang}}, \bibinfo {author} {\bibfnamefont
  {U.}~\bibnamefont {Patel}}, \bibinfo {author} {\bibfnamefont {Z.~L.}\
  \bibnamefont {Xiao}}, \ and\ \bibinfo {author} {\bibfnamefont {S.~Y.}\
  \bibnamefont {Li}},\ }\href@noop {} {\bibfield  {journal} {\bibinfo
  {journal} {Phys. Rev. B}\ }\textbf {\bibinfo {volume} {80}},\ \bibinfo
  {pages} {024518} (\bibinfo {year} {2009})}\BibitemShut {NoStop}%
\bibitem [{\citenamefont {Bourgeois-Hope}\ \emph {et~al.}(2016)\citenamefont
  {Bourgeois-Hope}, \citenamefont {Chi}, \citenamefont {Bonn}, \citenamefont
  {Liang}, \citenamefont {Hardy}, \citenamefont {Wolf}, \citenamefont
  {Meingast}, \citenamefont {Doiron-Leyraud},\ and\ \citenamefont
  {Taillefer}}]{hopearxiv}%
  \BibitemOpen
  \bibfield  {author} {\bibinfo {author} {\bibfnamefont {P.}~\bibnamefont
  {Bourgeois-Hope}}, \bibinfo {author} {\bibfnamefont {S.}~\bibnamefont {Chi}},
  \bibinfo {author} {\bibfnamefont {D.~A.}\ \bibnamefont {Bonn}}, \bibinfo
  {author} {\bibfnamefont {R.}~\bibnamefont {Liang}}, \bibinfo {author}
  {\bibfnamefont {W.~N.}\ \bibnamefont {Hardy}}, \bibinfo {author}
  {\bibfnamefont {T.}~\bibnamefont {Wolf}}, \bibinfo {author} {\bibfnamefont
  {C.}~\bibnamefont {Meingast}}, \bibinfo {author} {\bibfnamefont
  {N.}~\bibnamefont {Doiron-Leyraud}}, \ and\ \bibinfo {author} {\bibfnamefont
  {L.}~\bibnamefont {Taillefer}},\ }\href@noop {} {\bibfield  {journal}
  {\bibinfo  {journal} {Phys. Rev. Lett.}\ }\textbf {\bibinfo {volume} {117}},\
  \bibinfo {pages} {097003} (\bibinfo {year} {2016})}\BibitemShut {NoStop}%
\bibitem [{\citenamefont {Mazin}\ \emph {et~al.}(2008)\citenamefont {Mazin},
  \citenamefont {Singh}, \citenamefont {Johannes},\ and\ \citenamefont
  {Du}}]{MazinS}%
  \BibitemOpen
  \bibfield  {author} {\bibinfo {author} {\bibfnamefont {I.~I.}\ \bibnamefont
  {Mazin}}, \bibinfo {author} {\bibfnamefont {D.~J.}\ \bibnamefont {Singh}},
  \bibinfo {author} {\bibfnamefont {M.~D.}\ \bibnamefont {Johannes}}, \ and\
  \bibinfo {author} {\bibfnamefont {M.~H.}\ \bibnamefont {Du}},\ }\href@noop {}
  {\bibfield  {journal} {\bibinfo  {journal} {Phys. Rev. Lett.}\ }\textbf
  {\bibinfo {volume} {101}},\ \bibinfo {pages} {057003} (\bibinfo {year}
  {2008})}\BibitemShut {NoStop}%
\bibitem [{\citenamefont {Kontani}\ and\ \citenamefont
  {Onari}(2010)}]{KontaniPRL}%
  \BibitemOpen
  \bibfield  {author} {\bibinfo {author} {\bibfnamefont {H.}~\bibnamefont
  {Kontani}}\ and\ \bibinfo {author} {\bibfnamefont {S.}~\bibnamefont
  {Onari}},\ }\href@noop {} {\bibfield  {journal} {\bibinfo  {journal} {Phys.
  Rev. Lett.}\ }\textbf {\bibinfo {volume} {104}},\ \bibinfo {pages} {157001}
  (\bibinfo {year} {2010})}\BibitemShut {NoStop}%
\bibitem [{\citenamefont {Alloul}\ \emph {et~al.}(2009)\citenamefont {Alloul},
  \citenamefont {Bobroff}, \citenamefont {Gabay},\ and\ \citenamefont
  {Hirschfeld}}]{AlloulRevModPhys.81.45}%
  \BibitemOpen
  \bibfield  {author} {\bibinfo {author} {\bibfnamefont {H.}~\bibnamefont
  {Alloul}}, \bibinfo {author} {\bibfnamefont {J.}~\bibnamefont {Bobroff}},
  \bibinfo {author} {\bibfnamefont {M.}~\bibnamefont {Gabay}}, \ and\ \bibinfo
  {author} {\bibfnamefont {P.~J.}\ \bibnamefont {Hirschfeld}},\ }\href
  {\doibase 10.1103/RevModPhys.81.45} {\bibfield  {journal} {\bibinfo
  {journal} {Rev. Mod. Phys.}\ }\textbf {\bibinfo {volume} {81}},\ \bibinfo
  {pages} {45} (\bibinfo {year} {2009})}\BibitemShut {NoStop}%
\bibitem [{\citenamefont {Anderson}(1959)}]{Anderson}%
  \BibitemOpen
  \bibfield  {author} {\bibinfo {author} {\bibfnamefont {P.~W.}\ \bibnamefont
  {Anderson}},\ }\href@noop {} {\bibfield  {journal} {\bibinfo  {journal} {J.
  Phys. Chem. Solids}\ }\textbf {\bibinfo {volume} {11}},\ \bibinfo {pages}
  {26} (\bibinfo {year} {1959})}\BibitemShut {NoStop}%
\bibitem [{\citenamefont {Rullier-Albenque}\ \emph {et~al.}(2012)\citenamefont
  {Rullier-Albenque}, \citenamefont {Colson}, \citenamefont {Forget},\ and\
  \citenamefont {Alloul}}]{RullierPRL}%
  \BibitemOpen
  \bibfield  {author} {\bibinfo {author} {\bibfnamefont {F.}~\bibnamefont
  {Rullier-Albenque}}, \bibinfo {author} {\bibfnamefont {D.}~\bibnamefont
  {Colson}}, \bibinfo {author} {\bibfnamefont {A.}~\bibnamefont {Forget}}, \
  and\ \bibinfo {author} {\bibfnamefont {H.}~\bibnamefont {Alloul}},\
  }\href@noop {} {\bibfield  {journal} {\bibinfo  {journal} {Phys. Rev. Lett.}\
  }\textbf {\bibinfo {volume} {109}},\ \bibinfo {pages} {187005} (\bibinfo
  {year} {2012})}\BibitemShut {NoStop}%
\bibitem [{\citenamefont {Wang}\ \emph {et~al.}(2013)\citenamefont {Wang},
  \citenamefont {Kreisel}, \citenamefont {Hirschfeld},\ and\ \citenamefont
  {Mishra}}]{WangPRBdisorder}%
  \BibitemOpen
  \bibfield  {author} {\bibinfo {author} {\bibfnamefont {Y.}~\bibnamefont
  {Wang}}, \bibinfo {author} {\bibfnamefont {A.}~\bibnamefont {Kreisel}},
  \bibinfo {author} {\bibfnamefont {P.~J.}\ \bibnamefont {Hirschfeld}}, \ and\
  \bibinfo {author} {\bibfnamefont {V.}~\bibnamefont {Mishra}},\ }\href@noop {}
  {\bibfield  {journal} {\bibinfo  {journal} {Phys. Rev. B}\ }\textbf {\bibinfo
  {volume} {87}},\ \bibinfo {pages} {094504} (\bibinfo {year}
  {2013})}\BibitemShut {NoStop}%
\bibitem [{\citenamefont {Mizukami}\ \emph {et~al.}(2014)\citenamefont
  {Mizukami}, \citenamefont {Konczykowski}, \citenamefont {Kawamoto},
  \citenamefont {Kurata}, \citenamefont {Kasahara}, \citenamefont {Hashimoto},
  \citenamefont {Mishra}, \citenamefont {Kreisel}, \citenamefont {Wang},
  \citenamefont {Hirschfeld}, \citenamefont {Matsuda},\ and\ \citenamefont
  {Shibauchi}}]{MizukamiNatcom}%
  \BibitemOpen
  \bibfield  {author} {\bibinfo {author} {\bibfnamefont {Y.}~\bibnamefont
  {Mizukami}}, \bibinfo {author} {\bibfnamefont {M.}~\bibnamefont
  {Konczykowski}}, \bibinfo {author} {\bibfnamefont {Y.}~\bibnamefont
  {Kawamoto}}, \bibinfo {author} {\bibfnamefont {S.}~\bibnamefont {Kurata}},
  \bibinfo {author} {\bibfnamefont {S.}~\bibnamefont {Kasahara}}, \bibinfo
  {author} {\bibfnamefont {K.}~\bibnamefont {Hashimoto}}, \bibinfo {author}
  {\bibfnamefont {V.}~\bibnamefont {Mishra}}, \bibinfo {author} {\bibfnamefont
  {A.}~\bibnamefont {Kreisel}}, \bibinfo {author} {\bibfnamefont
  {Y.}~\bibnamefont {Wang}}, \bibinfo {author} {\bibfnamefont {P.~J.}\
  \bibnamefont {Hirschfeld}}, \bibinfo {author} {\bibfnamefont
  {Y.}~\bibnamefont {Matsuda}}, \ and\ \bibinfo {author} {\bibfnamefont
  {T.}~\bibnamefont {Shibauchi}},\ }\href@noop {} {\bibfield  {journal}
  {\bibinfo  {journal} {Nat. Commun.}\ }\textbf {\bibinfo {volume} {5}},\
  \bibinfo {pages} {5657} (\bibinfo {year} {2014})}\BibitemShut {NoStop}%
\bibitem [{\citenamefont {Onari}\ and\ \citenamefont
  {Kontani}(2009)}]{OnariPRLdisorder}%
  \BibitemOpen
  \bibfield  {author} {\bibinfo {author} {\bibfnamefont {S.}~\bibnamefont
  {Onari}}\ and\ \bibinfo {author} {\bibfnamefont {H.}~\bibnamefont
  {Kontani}},\ }\href@noop {} {\bibfield  {journal} {\bibinfo  {journal} {Phys.
  Rev. Lett.}\ }\textbf {\bibinfo {volume} {103}},\ \bibinfo {pages} {177001}
  (\bibinfo {year} {2009})}\BibitemShut {NoStop}%
\bibitem [{\citenamefont {Prozorov}\ \emph {et~al.}(2014)\citenamefont
  {Prozorov}, \citenamefont {Konczykowski}, \citenamefont {Tanatar},
  \citenamefont {Thaler}, \citenamefont {Bud’ko}, \citenamefont {Canfield},
  \citenamefont {Mishra},\ and\ \citenamefont
  {Hirschfeld}}]{ProzorovPRXBaPirra}%
  \BibitemOpen
  \bibfield  {author} {\bibinfo {author} {\bibfnamefont {R.}~\bibnamefont
  {Prozorov}}, \bibinfo {author} {\bibfnamefont {M.}~\bibnamefont
  {Konczykowski}}, \bibinfo {author} {\bibfnamefont {M.~A.}\ \bibnamefont
  {Tanatar}}, \bibinfo {author} {\bibfnamefont {A.}~\bibnamefont {Thaler}},
  \bibinfo {author} {\bibfnamefont {S.~L.}\ \bibnamefont {Bud’ko}}, \bibinfo
  {author} {\bibfnamefont {P.~C.}\ \bibnamefont {Canfield}}, \bibinfo {author}
  {\bibfnamefont {V.}~\bibnamefont {Mishra}}, \ and\ \bibinfo {author}
  {\bibfnamefont {P.~J.}\ \bibnamefont {Hirschfeld}},\ }\href@noop {}
  {\bibfield  {journal} {\bibinfo  {journal} {Phys. Rev. X}\ }\textbf {\bibinfo
  {volume} {4}},\ \bibinfo {pages} {041032} (\bibinfo {year}
  {2014})}\BibitemShut {NoStop}%
\bibitem [{\citenamefont {Sun}\ \emph {et~al.}(2016{\natexlab{a}})\citenamefont
  {Sun}, \citenamefont {Pyon},\ and\ \citenamefont
  {Tamegai}}]{SunPhysRevB.93.104502}%
  \BibitemOpen
  \bibfield  {author} {\bibinfo {author} {\bibfnamefont {Y.}~\bibnamefont
  {Sun}}, \bibinfo {author} {\bibfnamefont {S.}~\bibnamefont {Pyon}}, \ and\
  \bibinfo {author} {\bibfnamefont {T.}~\bibnamefont {Tamegai}},\ }\href
  {\doibase 10.1103/PhysRevB.93.104502} {\bibfield  {journal} {\bibinfo
  {journal} {Phys. Rev. B}\ }\textbf {\bibinfo {volume} {93}},\ \bibinfo
  {pages} {104502} (\bibinfo {year} {2016}{\natexlab{a}})}\BibitemShut
  {NoStop}%
\bibitem [{\citenamefont {Sun}\ \emph {et~al.}(2016{\natexlab{b}})\citenamefont
  {Sun}, \citenamefont {Yamada}, \citenamefont {Pyon},\ and\ \citenamefont
  {Tamegai}}]{SunYueAMRFeSe}%
  \BibitemOpen
  \bibfield  {author} {\bibinfo {author} {\bibfnamefont {Y.}~\bibnamefont
  {Sun}}, \bibinfo {author} {\bibfnamefont {T.}~\bibnamefont {Yamada}},
  \bibinfo {author} {\bibfnamefont {S.}~\bibnamefont {Pyon}}, \ and\ \bibinfo
  {author} {\bibfnamefont {T.}~\bibnamefont {Tamegai}},\ }\href@noop {}
  {\bibfield  {journal} {\bibinfo  {journal} {Phys. Rev. B}\ }\textbf {\bibinfo
  {volume} {94}},\ \bibinfo {pages} {134505} (\bibinfo {year}
  {2016}{\natexlab{b}})}\BibitemShut {NoStop}%
\bibitem [{\citenamefont {Ziegler}\ \emph {et~al.}(1985)\citenamefont
  {Ziegler}, \citenamefont {Biersack},\ and\ \citenamefont
  {Littmark}}]{irradiationrange}%
  \BibitemOpen
  \bibfield  {author} {\bibinfo {author} {\bibfnamefont {J.}~\bibnamefont
  {Ziegler}}, \bibinfo {author} {\bibfnamefont {J.}~\bibnamefont {Biersack}}, \
  and\ \bibinfo {author} {\bibfnamefont {U.}~\bibnamefont {Littmark}},\
  }\href@noop {} {\emph {\bibinfo {title} {The Stopping and Range of Ions in
  Solids}}}\ (\bibinfo  {publisher} {New York: Pergamon},\ \bibinfo {year}
  {1985})\BibitemShut {NoStop}%
\bibitem [{sup()}]{supplement}%
  \BibitemOpen
  \href@noop {} {\ }\bibinfo {note} {{See Supplemental Material at
  [http://link.aps.org/supplemental/10.1103/PhysRevB.96. 140505] for the
  confirmations of no strain effect or annealing effect or changing on the
  fundamental electronic structure after H$^+$-irradiation, the definition of
  $T_c$, and more discussion about the defects introduced by
  irradiation}}\BibitemShut {NoStop}%
\bibitem [{\citenamefont {Bouquet}\ \emph {et~al.}(2002)\citenamefont
  {Bouquet}, \citenamefont {Wang}, \citenamefont {Sheikin}, \citenamefont
  {Plackowski}, \citenamefont {Junod}, \citenamefont {Lee},\ and\ \citenamefont
  {Tajima}}]{MgB2PhysRevLetttwogap}%
  \BibitemOpen
  \bibfield  {author} {\bibinfo {author} {\bibfnamefont {F.}~\bibnamefont
  {Bouquet}}, \bibinfo {author} {\bibfnamefont {Y.}~\bibnamefont {Wang}},
  \bibinfo {author} {\bibfnamefont {I.}~\bibnamefont {Sheikin}}, \bibinfo
  {author} {\bibfnamefont {T.}~\bibnamefont {Plackowski}}, \bibinfo {author}
  {\bibfnamefont {A.}~\bibnamefont {Junod}}, \bibinfo {author} {\bibfnamefont
  {S.}~\bibnamefont {Lee}}, \ and\ \bibinfo {author} {\bibfnamefont
  {S.}~\bibnamefont {Tajima}},\ }\href@noop {} {\bibfield  {journal} {\bibinfo
  {journal} {Phys. Rev. Lett.}\ }\textbf {\bibinfo {volume} {89}},\ \bibinfo
  {pages} {257001} (\bibinfo {year} {2002})}\BibitemShut {NoStop}%
\bibitem [{\citenamefont {Nakajima}\ \emph {et~al.}(2010)\citenamefont
  {Nakajima}, \citenamefont {Taen}, \citenamefont {Tsuchiya}, \citenamefont
  {Tamegai}, \citenamefont {Kitamura},\ and\ \citenamefont
  {Murakami}}]{NakajimaPRBBaCoirra}%
  \BibitemOpen
  \bibfield  {author} {\bibinfo {author} {\bibfnamefont {Y.}~\bibnamefont
  {Nakajima}}, \bibinfo {author} {\bibfnamefont {T.}~\bibnamefont {Taen}},
  \bibinfo {author} {\bibfnamefont {Y.}~\bibnamefont {Tsuchiya}}, \bibinfo
  {author} {\bibfnamefont {T.}~\bibnamefont {Tamegai}}, \bibinfo {author}
  {\bibfnamefont {H.}~\bibnamefont {Kitamura}}, \ and\ \bibinfo {author}
  {\bibfnamefont {T.}~\bibnamefont {Murakami}},\ }\href@noop {} {\bibfield
  {journal} {\bibinfo  {journal} {Phys. Rev. B}\ }\textbf {\bibinfo {volume}
  {82}},\ \bibinfo {pages} {220504} (\bibinfo {year} {2010})}\BibitemShut
  {NoStop}%
\bibitem [{\citenamefont {Taen}\ \emph {et~al.}(2013)\citenamefont {Taen},
  \citenamefont {Ohtake}, \citenamefont {Akiyama}, \citenamefont {Inoue},
  \citenamefont {Sun}, \citenamefont {Pyon}, \citenamefont {Tamegai},\ and\
  \citenamefont {Kitamura}}]{TaenPRBBaKirr}%
  \BibitemOpen
  \bibfield  {author} {\bibinfo {author} {\bibfnamefont {T.}~\bibnamefont
  {Taen}}, \bibinfo {author} {\bibfnamefont {F.}~\bibnamefont {Ohtake}},
  \bibinfo {author} {\bibfnamefont {H.}~\bibnamefont {Akiyama}}, \bibinfo
  {author} {\bibfnamefont {H.}~\bibnamefont {Inoue}}, \bibinfo {author}
  {\bibfnamefont {Y.}~\bibnamefont {Sun}}, \bibinfo {author} {\bibfnamefont
  {S.}~\bibnamefont {Pyon}}, \bibinfo {author} {\bibfnamefont {T.}~\bibnamefont
  {Tamegai}}, \ and\ \bibinfo {author} {\bibfnamefont {H.}~\bibnamefont
  {Kitamura}},\ }\href@noop {} {\bibfield  {journal} {\bibinfo  {journal}
  {Phys. Rev. B}\ }\textbf {\bibinfo {volume} {88}},\ \bibinfo {pages} {224514}
  (\bibinfo {year} {2013})}\BibitemShut {NoStop}%
\bibitem [{\citenamefont {Kreisel}\ \emph {et~al.}(2015)\citenamefont
  {Kreisel}, \citenamefont {Mukherjee}, \citenamefont {Hirschfeld},\ and\
  \citenamefont {Andersen}}]{KreiselPRB}%
  \BibitemOpen
  \bibfield  {author} {\bibinfo {author} {\bibfnamefont {A.}~\bibnamefont
  {Kreisel}}, \bibinfo {author} {\bibfnamefont {S.}~\bibnamefont {Mukherjee}},
  \bibinfo {author} {\bibfnamefont {P.~J.}\ \bibnamefont {Hirschfeld}}, \ and\
  \bibinfo {author} {\bibfnamefont {B.~M.}\ \bibnamefont {Andersen}},\
  }\href@noop {} {\bibfield  {journal} {\bibinfo  {journal} {Phys. Rev. B}\
  }\textbf {\bibinfo {volume} {92}},\ \bibinfo {pages} {224515} (\bibinfo
  {year} {2015})}\BibitemShut {NoStop}%
\bibitem [{\citenamefont {Watson}\ \emph {et~al.}(2015)\citenamefont {Watson},
  \citenamefont {Kim}, \citenamefont {Haghighirad}, \citenamefont {Davies},
  \citenamefont {McCollam}, \citenamefont {Narayanan}, \citenamefont {Blake},
  \citenamefont {Chen}, \citenamefont {Ghannadzadeh}, \citenamefont
  {Schofield}, \citenamefont {Hoesch}, \citenamefont {Meingast}, \citenamefont
  {Wolf},\ and\ \citenamefont {Coldea}}]{WatsonPRB92}%
  \BibitemOpen
  \bibfield  {author} {\bibinfo {author} {\bibfnamefont {M.~D.}\ \bibnamefont
  {Watson}}, \bibinfo {author} {\bibfnamefont {T.~K.}\ \bibnamefont {Kim}},
  \bibinfo {author} {\bibfnamefont {A.~A.}\ \bibnamefont {Haghighirad}},
  \bibinfo {author} {\bibfnamefont {N.~R.}\ \bibnamefont {Davies}}, \bibinfo
  {author} {\bibfnamefont {A.}~\bibnamefont {McCollam}}, \bibinfo {author}
  {\bibfnamefont {A.}~\bibnamefont {Narayanan}}, \bibinfo {author}
  {\bibfnamefont {S.~F.}\ \bibnamefont {Blake}}, \bibinfo {author}
  {\bibfnamefont {Y.~L.}\ \bibnamefont {Chen}}, \bibinfo {author}
  {\bibfnamefont {S.}~\bibnamefont {Ghannadzadeh}}, \bibinfo {author}
  {\bibfnamefont {A.~J.}\ \bibnamefont {Schofield}}, \bibinfo {author}
  {\bibfnamefont {M.}~\bibnamefont {Hoesch}}, \bibinfo {author} {\bibfnamefont
  {C.}~\bibnamefont {Meingast}}, \bibinfo {author} {\bibfnamefont
  {T.}~\bibnamefont {Wolf}}, \ and\ \bibinfo {author} {\bibfnamefont {A.~I.}\
  \bibnamefont {Coldea}},\ }\href@noop {} {\bibfield  {journal} {\bibinfo
  {journal} {Phys. Rev. B}\ }\textbf {\bibinfo {volume} {91}},\ \bibinfo
  {pages} {155106} (\bibinfo {year} {2015})}\BibitemShut {NoStop}%
\bibitem [{\citenamefont {Fan}\ \emph {et~al.}(2015)\citenamefont {Fan},
  \citenamefont {Zhang}, \citenamefont {Liu}, \citenamefont {Yan},
  \citenamefont {Ren}, \citenamefont {Peng}, \citenamefont {Xu}, \citenamefont
  {Xie}, \citenamefont {Hu}, \citenamefont {Zhang},\ and\ \citenamefont
  {Feng}}]{FanNatPhyMonoFeSe}%
  \BibitemOpen
  \bibfield  {author} {\bibinfo {author} {\bibfnamefont {Q.}~\bibnamefont
  {Fan}}, \bibinfo {author} {\bibfnamefont {W.~H.}\ \bibnamefont {Zhang}},
  \bibinfo {author} {\bibfnamefont {X.}~\bibnamefont {Liu}}, \bibinfo {author}
  {\bibfnamefont {Y.~J.}\ \bibnamefont {Yan}}, \bibinfo {author} {\bibfnamefont
  {M.~Q.}\ \bibnamefont {Ren}}, \bibinfo {author} {\bibfnamefont
  {R.}~\bibnamefont {Peng}}, \bibinfo {author} {\bibfnamefont {H.~C.}\
  \bibnamefont {Xu}}, \bibinfo {author} {\bibfnamefont {B.~P.}\ \bibnamefont
  {Xie}}, \bibinfo {author} {\bibfnamefont {J.~P.}\ \bibnamefont {Hu}},
  \bibinfo {author} {\bibfnamefont {T.}~\bibnamefont {Zhang}}, \ and\ \bibinfo
  {author} {\bibfnamefont {D.~L.}\ \bibnamefont {Feng}},\ }\href@noop {}
  {\bibfield  {journal} {\bibinfo  {journal} {Nat Phys}\ }\textbf {\bibinfo
  {volume} {11}},\ \bibinfo {pages} {946} (\bibinfo {year} {2015})}\BibitemShut
  {NoStop}%
\bibitem [{\citenamefont {Teknowijoyo}\ \emph {et~al.}(2016)\citenamefont
  {Teknowijoyo}, \citenamefont {Cho}, \citenamefont {Tanatar}, \citenamefont
  {Gonzales}, \citenamefont {B\"{o}hmer}, \citenamefont {Cavani}, \citenamefont
  {Mishra}, \citenamefont {Hirschfeld}, \citenamefont {Bud'ko}, \citenamefont
  {Canfield},\ and\ \citenamefont {Prozorov}}]{Teknowijoyoarxiv}%
  \BibitemOpen
  \bibfield  {author} {\bibinfo {author} {\bibfnamefont {S.}~\bibnamefont
  {Teknowijoyo}}, \bibinfo {author} {\bibfnamefont {K.}~\bibnamefont {Cho}},
  \bibinfo {author} {\bibfnamefont {M.~A.}\ \bibnamefont {Tanatar}}, \bibinfo
  {author} {\bibfnamefont {J.}~\bibnamefont {Gonzales}}, \bibinfo {author}
  {\bibfnamefont {A.~E.}\ \bibnamefont {B\"{o}hmer}}, \bibinfo {author}
  {\bibfnamefont {O.}~\bibnamefont {Cavani}}, \bibinfo {author} {\bibfnamefont
  {V.}~\bibnamefont {Mishra}}, \bibinfo {author} {\bibfnamefont {P.~J.}\
  \bibnamefont {Hirschfeld}}, \bibinfo {author} {\bibfnamefont {S.~L.}\
  \bibnamefont {Bud'ko}}, \bibinfo {author} {\bibfnamefont {P.~C.}\
  \bibnamefont {Canfield}}, \ and\ \bibinfo {author} {\bibfnamefont
  {R.}~\bibnamefont {Prozorov}},\ }\href@noop {} {\bibfield  {journal}
  {\bibinfo  {journal} {Phys. Rev. B}\ }\textbf {\bibinfo {volume} {94}},\
  \bibinfo {pages} {064521} (\bibinfo {year} {2016})}\BibitemShut {NoStop}%
\bibitem [{\citenamefont {Smylie}\ \emph {et~al.}(2016)\citenamefont {Smylie},
  \citenamefont {Leroux}, \citenamefont {Mishra}, \citenamefont {Fang},
  \citenamefont {Taddei}, \citenamefont {Chmaissem}, \citenamefont {Claus},
  \citenamefont {Kayani}, \citenamefont {Snezhko}, \citenamefont {Welp},\ and\
  \citenamefont {Kwok}}]{SmyliePhysRevB}%
  \BibitemOpen
  \bibfield  {author} {\bibinfo {author} {\bibfnamefont {M.~P.}\ \bibnamefont
  {Smylie}}, \bibinfo {author} {\bibfnamefont {M.}~\bibnamefont {Leroux}},
  \bibinfo {author} {\bibfnamefont {V.}~\bibnamefont {Mishra}}, \bibinfo
  {author} {\bibfnamefont {L.}~\bibnamefont {Fang}}, \bibinfo {author}
  {\bibfnamefont {K.~M.}\ \bibnamefont {Taddei}}, \bibinfo {author}
  {\bibfnamefont {O.}~\bibnamefont {Chmaissem}}, \bibinfo {author}
  {\bibfnamefont {H.}~\bibnamefont {Claus}}, \bibinfo {author} {\bibfnamefont
  {A.}~\bibnamefont {Kayani}}, \bibinfo {author} {\bibfnamefont
  {A.}~\bibnamefont {Snezhko}}, \bibinfo {author} {\bibfnamefont
  {U.}~\bibnamefont {Welp}}, \ and\ \bibinfo {author} {\bibfnamefont {W.-K.}\
  \bibnamefont {Kwok}},\ }\href {\doibase 10.1103/PhysRevB.93.115119}
  {\bibfield  {journal} {\bibinfo  {journal} {Phys. Rev. B}\ }\textbf {\bibinfo
  {volume} {93}},\ \bibinfo {pages} {115119} (\bibinfo {year}
  {2016})}\BibitemShut {NoStop}%
\bibitem [{\citenamefont {Watashige}\ \emph {et~al.}(2015)\citenamefont
  {Watashige}, \citenamefont {Tsutsumi}, \citenamefont {Hanaguri},
  \citenamefont {Kohsaka}, \citenamefont {Kasahara}, \citenamefont {Furusaki},
  \citenamefont {Sigrist}, \citenamefont {Meingast}, \citenamefont {Wolf},
  \citenamefont {Löhneysen}, \citenamefont {Shibauchi},\ and\ \citenamefont
  {Matsuda}}]{WatashigePRX}%
  \BibitemOpen
  \bibfield  {author} {\bibinfo {author} {\bibfnamefont {T.}~\bibnamefont
  {Watashige}}, \bibinfo {author} {\bibfnamefont {Y.}~\bibnamefont {Tsutsumi}},
  \bibinfo {author} {\bibfnamefont {T.}~\bibnamefont {Hanaguri}}, \bibinfo
  {author} {\bibfnamefont {Y.}~\bibnamefont {Kohsaka}}, \bibinfo {author}
  {\bibfnamefont {S.}~\bibnamefont {Kasahara}}, \bibinfo {author}
  {\bibfnamefont {A.}~\bibnamefont {Furusaki}}, \bibinfo {author}
  {\bibfnamefont {M.}~\bibnamefont {Sigrist}}, \bibinfo {author} {\bibfnamefont
  {C.}~\bibnamefont {Meingast}}, \bibinfo {author} {\bibfnamefont
  {T.}~\bibnamefont {Wolf}}, \bibinfo {author} {\bibfnamefont {H.~v.}\
  \bibnamefont {Löhneysen}}, \bibinfo {author} {\bibfnamefont
  {T.}~\bibnamefont {Shibauchi}}, \ and\ \bibinfo {author} {\bibfnamefont
  {Y.}~\bibnamefont {Matsuda}},\ }\href@noop {} {\bibfield  {journal} {\bibinfo
   {journal} {Phys. Rev. X}\ }\textbf {\bibinfo {volume} {5}},\ \bibinfo
  {pages} {031022} (\bibinfo {year} {2015})}\BibitemShut {NoStop}%
\bibitem [{\citenamefont {Sprau}\ \emph {et~al.}(2017)\citenamefont {Sprau},
  \citenamefont {Kostin}, \citenamefont {Kreisel}, \citenamefont {B{\"o}hmer},
  \citenamefont {Taufour}, \citenamefont {Canfield}, \citenamefont {Mukherjee},
  \citenamefont {Hirschfeld}, \citenamefont {Andersen},\ and\ \citenamefont
  {Davis}}]{BQPIScience}%
  \BibitemOpen
  \bibfield  {author} {\bibinfo {author} {\bibfnamefont {P.~O.}\ \bibnamefont
  {Sprau}}, \bibinfo {author} {\bibfnamefont {A.}~\bibnamefont {Kostin}},
  \bibinfo {author} {\bibfnamefont {A.}~\bibnamefont {Kreisel}}, \bibinfo
  {author} {\bibfnamefont {A.~E.}\ \bibnamefont {B{\"o}hmer}}, \bibinfo
  {author} {\bibfnamefont {V.}~\bibnamefont {Taufour}}, \bibinfo {author}
  {\bibfnamefont {P.~C.}\ \bibnamefont {Canfield}}, \bibinfo {author}
  {\bibfnamefont {S.}~\bibnamefont {Mukherjee}}, \bibinfo {author}
  {\bibfnamefont {P.~J.}\ \bibnamefont {Hirschfeld}}, \bibinfo {author}
  {\bibfnamefont {B.~M.}\ \bibnamefont {Andersen}}, \ and\ \bibinfo {author}
  {\bibfnamefont {J.~C.~S.}\ \bibnamefont {Davis}},\ }\href@noop {} {\bibfield
  {journal} {\bibinfo  {journal} {Science}\ }\textbf {\bibinfo {volume}
  {357}},\ \bibinfo {pages} {75} (\bibinfo {year} {2017})}\BibitemShut
  {NoStop}%
\end{thebibliography}%

\end{document}